\documentclass[review]{elsarticle}

\def\msun{M_\odot}
\def\Sgr{{ SgrA* }}

\begin{document}

\begin{frontmatter}

\title{Our Supermassive Black Hole Rivaled the Sun\\ in the Ancient X-ray Sky}

\author{Pau Amaro-Seoane}
\address{
Institut de Ci{\`e}ncies de l'Espai (CSIC-IEEC) at Campus UAB,\\ 
Carrer de Can Magrans s/n 08193 Barcelona, Spain\\
Kavli Institute for Astronomy and Astrophysics, Beijing 100871, China\\
Institute of Applied Mathematics, Academy of Mathematics and Systems Science,\\
CAS, Beijing 100190, China\\
Zentrum f{\"u}r Astronomie und Astrophysik, TU Berlin,\\ 
Hardenbergstra{\ss}e 36, 10623 Berlin, Germany
}

\author{Xian Chen\fnref{myfootnote}}
\address{
Astronomy Department, School of Physics, Peking University, 100871 Beijing, China\\
Kavli Institute for Astronomy and Astrophysics, Beijing 100871, China
}
\fntext[myfootnote]{Corresponding author.}


\begin{abstract}

Sagittarius A* (SgrA*) lying in the Galactic Centre $8$ kpc from Earth, hosts
	the closest supermassive black hole known to us. It is now inactive,
	but there is evidence indicating that about six million years ago it
	underwent a powerful outburst where the luminosity could have
	approached the Eddington limit.  Motivated by the fact that in
	extragalaxies the supermassive black holes with similar masses and
	near-Eddington luminosities are usually strong X-ray emitters, we
	calculate here the X-ray luminosity of SgrA*.  For that, we assume that
	the outburst was due to accretion of gas or the tidal disruption of a
	star.  We show that these cases could precipitate on Earth a hard X-ray
	(i.e. $h\nu>2~{\rm keV}$) flux comparable to that from the current
	quiescent sun.  The flux in harder energy band $20~{\rm
	keV}<h\nu<100~{\rm keV}$, however, surpasses that from an X-class solar
	flare, and the irradiation timescale is also much longer, ranging from
	weeks to $10^5$ years depending on the outburst scenario.  In the solar
	system gas giants will suffer the biggest impact in their atmospheres.
	Lower-mass planets such as Earth receive a level of radiation that
	might have played a role in the evolution of their primitive
	atmospheres, so that a detailed study of the consequences deserves
	further investigation.  Planetary systems closer to SgrA* receive
	higher irradiance levels, making them more likely uninhabitable.

\end{abstract}



\end{frontmatter}

%\linenumbers

\section{Introduction}

Supermassive black holes (SMBHs) intermittently undergo electromagnetic
outbursts at the expense of the gravitational energies of the infalling gas or
stars, but during most of their lifetime they are quiescent, ``inactive'', and
are found in the nuclei of galaxies \citep{kormendy13}. Our own Galaxy harbours
one, SgrA*, with a mass of about $4\times10^6\,M_{\odot}$ \citep[see the review
of][ and references therein]{genzel10}.

Although it is accepted that at early ages our Milky Way (MW) was an Active
Galactic Nucleus (AGN), the possibility that during the past few Myrs SgrA* had
an activity in the range of an AGN has not been addressed because, at least
partly, AGNs have been thought to be uncommon. The common believe was that AGNs
are mostly triggered via galaxy mergers. There is evidence suggesting that our
Galaxy has not experienced a major merger during the past $10$ billion years
\citep{gilmore02}. This led to believe that the MW could not have had such an
activity recently, because the merger timescale is considered to be too long.

However, this situation has changed significantly. Recent studies of AGNs in
the local universe suggest that SMBHs less massive than $10^7~\msun$, like
SgrA*, could be frequently activated by stochastic processes such as
gravitational instability \citep{hopkins06}. The frequency could be as high as
once every $10^7-10^8$ years, with each burst lasting about $10^5$ years \citep{hopkins06}.
Because of them, it is not ruled out that the Galactic Centre could have had a
significant activity at some point during the past $10$ Myrs.

On the observational front, there is (mounting) observational evidence that
\Sgr could have undergone an energetic outburst about $2-8$ Myr ago
\citep{MezgerEtAl1996,nay05,BlandHawthorn2013,Amaro-SeoaneChen2014,ChenAmaro-Seoane2014a,ChenAmaro-Seoane2014b},
strengthened by the discovery of the so-called ``Fermi bubbles''. These are two
gamma-ray emitting bubbles located below and above the GC and extending as much
as 10~kpc \citep{DoblerEtAl2010,su10} which can be explained either in terms of
AGN outflows \citep{ZubovasEtAl2011,GuoMathews2012,MouEtAl2014}, star capture
events in the last $\sim\,10$ Myrs \citep{ChengEtAl2011} or a nuclear
starburst \citep{su10}. 
The observed gamma-ray activity fits the picture that
until very recently SgrA* was orders of magnitude more luminous in X-ray than
it currently is \citep{PontiEtAl2013}. If the bubble is caused by an AGN,
the kinematics is consistent with an outburst which lasts $0.1-1$
Myr \cite{ZubovasEtAl2011}.

It is difficult to infer the corresponding bolometric luminosity, and the
current estimation of the peak luminosity falls in the range of about
$(3-100)\%$ of the Eddington limit $L_{\rm Edd}\simeq5\times10^{44}~{\rm
erg~s^{-1}}$ \citep{nay05,BlandHawthorn13}, which is a usual value for an AGN.
This level of activity fits the framework of the aforementioned stochastic AGN
model.

Current observations show that AGNs radiating at a luminosity close to the
Eddington limit emit about $1\%$ of their bolometric luminosity in hard X-ray
($h\nu>2$ keV) \citep{jin12III,vasudevan14}. Theoretically, we can explain this
in the context of a hot corona screening the accretion disc, as put forward by
\citep{lu99,wang04,QiaoLiu2017}. This means that the past activity of the MW
must have had a relatively important hard X-ray component.

The implications are interesting because of the potential impact that this hard
X-ray could have had on Earth in the past. We note that this idea is not new.
Indeed, \cite{clarke81,laviolette87,gonzalez05} looked into this possibility
but did not take into account that this might have happened much more recently
than they thought, because they lacked the observational evidence and
theoretical framework.

We note that the consequences of an enhanced hard X-ray flux have been studied
in detail in a different scenario: that of a near-by gamma-ray burst
\citep{dartnell11,melott11}. The authors claimed that a sudden enhancement of
this irradiation on Earth could increase the ionization of the upper
atmosphere, leading to a chain of secondary reactions, among which ozone
depletion, to a degree which depends on the fluence.

The magnitude of the impact in this or any other scenario is determined by the
intensity and duration of the X-ray irradiation, as well as the spectral shape
of the ionizing radiation. In this article we address these factors in the
scenario that we propose: A possible recent active past of \Sgr.

\section{X-ray irradiation from Sagittarius A*}

In this section we consider two different possibilities for the outburst of
radiation originating at the Galactic Centre: A relatively recent AGN-like
episode and a stellar tidal disruption event in a  quiescent \Sgr. We compare
the X-ray irradiation levels in both scenarios with those originating from
three degrees of current solar activity: a maximum, a minimum and a strong
solar flare.

\subsection{A recent AGN-like episode}\label{sec:AGNSED}

We can derive the X-ray irradiation that potentially hit Earth in the past
by using the model of \cite{jin12III} for a sample of unobscured AGNs.
The unabsorbed, or unobscured, X-ray spectral energy distribution (SED,
i.e. the luminosity per logarithmic energy bin) of an AGN can be modeled
with a power law, which is consistent with the corona model, as discussed
in the introduction.

The model of \cite{jin12III} provides us with the normalization and power-law index.
According to these templates, given the ratio $\lambda_{\rm Edd}\equiv L_{\rm bol}/L_{\rm Edd}$
between the bolometric luminosity $L_{\rm bol}$ and the Eddington luminosity
$L_{\rm Edd}$, the luminosity $L_{\rm 2-10~keV}$ of the photons in the band
$2~{\rm keV}<h\nu<10~{\rm keV}$ scales as

\begin{equation}
\log{(L_{\rm 2-10~keV}/L_{\rm
bol})}=-0.773\log(\lambda_{\rm Edd})-2.004.
\end{equation}

\noindent
The photon index $\Gamma$, the index of a power law which characterizes the
number of photons in each energy bin, is given by

\begin{equation}
\Gamma=0.564\log(\lambda_{\rm Edd})+2.246.
\end{equation}

\noindent
The SED will have a power-law index of $-\Gamma+2$. When converting the SED to the
spectral irradiance on Earth ($\nu F_\nu$), we apply the extinction law of the Galactic
plane ($A_K=7.0$ and $R_V=3.1$), following \cite{tan04}. We do not consider
the radiation from the accretion disc because, given the Eddington
ratio of our interest ($\lambda_{\rm Edd}>0.1$), the disc SED peaks at UV and
soft X-ray bands, where the photons will be completely absorbed due to the high
extinction. I.e. the irradiation comes mostly from the corona and we neglect the
disc.

\begin{figure}
\includegraphics[width=84mm]{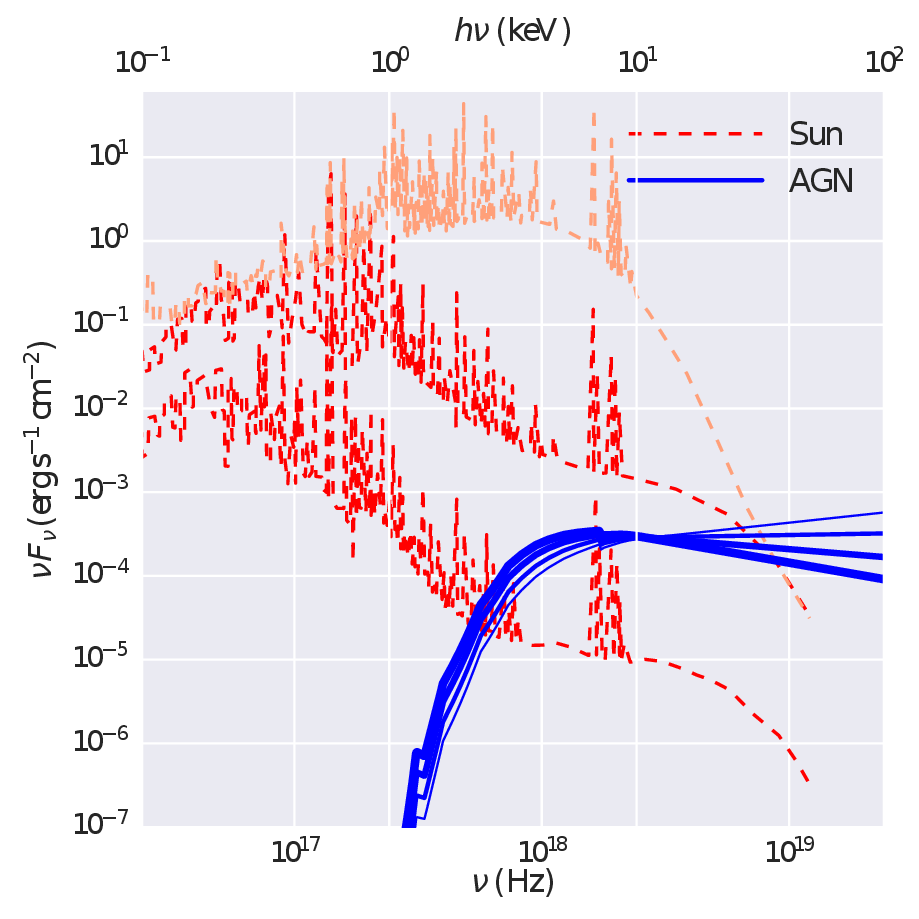}
\caption{
Comparison of X-ray irradiance on Earth from the Sun today (the dashed lines) and a potential
AGN-like episode (the four solid lines) originating at the Galactic Centre. The
thickness of the solid lines corresponds to different accretion rates, more
specifically to $\lambda_{\rm Edd}=(0.1,\,0.3,\,1,\,3)$, respectively (thicker means more efficient
accretion). The dashed lines, taken from \cite{peres00}, from the bottom to the top
correspond to a minimum irradiance during the solar cycle, a maximum one,
and the top-most corresponds to an intense X-class solar flare.
\label{fig1}
}
\end{figure}

Having derived the X-ray flux as a function of photon energy on Earth
originating from a past violent outburst in\Sgr, we compare it to current
levels from the Sun, as provided by \cite{peres00}, in their Figure 6. We
depict the comparison in Figure~\ref{fig1}.

We compare our analytical fitting with detailed, numerical simulations of
an accretion disc including the corona structure.
The numerical simulations follow the model of \cite{LiuEtAl2016}. We adopt
different values of
the ratio of the sum of gas pressure and radiation pressure to the magnetic pressure in the disc,
$\beta$, of $50$ and $100$, and different values for the
$\alpha$ viscosity of the disc, of $0.1$ and $0.3$. For the range of accretion values that we use in this
work, we find no significant difference between the empirical flux and the numerical simulations, so that
for convenience we use the former.

The most important features from Figure~\ref{fig1} are that: (i) At about $h\nu>2$ keV the irradiance
originating from an active past of \Sgr exceeds that from the Sun during the solar minimum,
regardless of $\lambda_{\rm Edd}$. (ii) Starting at about $20-30$ keV, the SED
from the assumption that \Sgr was active in the past starts to become higher than that from the
highest value which we can expect from the Sun, a strong X-class solar flare. (iii) At around
$h\nu>10$ keV the relation between irradiance and $\lambda_{\rm Edd}$ changes, meaning that
lower values of accretion result in an even higher irradiance. This has already been explained in the
disc corona model \citep{lu99,wang04}. (iv) In the assumption that \Sgr was active in the past,
the SED extends out to higher frequencies than what we can expect from the irrandiance of the Sun,
up to $h\nu\sim10^2$ keV \citep{vasudevan14}, the band of soft $\gamma$-ray. This makes the
potential active episode of \Sgr brighter than the Sun in the hard-X-ray/soft-$\gamma$-ray, for all possible solar scenarios.

\subsection{A tidal disruption event}

The scenario that we consider now is the following: The X-ray luminosity of \Sgr can
be temporarily enhanced by a tidal disruption of a star close to the SMBH \citep{rees88,komossa12}. The event rate is of about once every $10^4-10^5$ years during the quiescent phase of
\Sgr \citep{merritt10}.

Historically, we know that TDEs must have a (1) soft, thermal SED with an effective temperature of $kT\simeq0.1$ keV, as already put forward by the pioneering theoretical work of \cite{rees88}, but see also \cite{ulmer99,li02,chen13} for more recent works. Indeed, this soft component has been observed in TDE candidates \citep{komossa12}.

However, as we started harvesting more and more data, observations have revealed (2) a hard X-ray emission from some TDE candidates \citep{lin11,saxton12,niko13}. These TDE candidates exhibit strong power-law components with $2<\Gamma<5$ in the energy band $h\nu>2$ keV. This hard power-law component
has an exponent of $\Gamma=3$ whose luminosity is $10\%$ of the black-body emission, as empirically fitted by \cite{strubbe11}

A possible theoretical explanation could be that these originate from
shock-heated, incompletely-thermalized material around the SMBH \citep{ulmer99,strubbe11}. In these models, the emission would be isotropic.
We note here that we have observed hard X-ray irradiances exceeding by orders
of magnitude the maximum what can be expected from these models
\citep{BurrowsEtAl2011,LevanEtAl2011,CenkoEtAl2012,BrownEtAl2015}.  In these
cases, a highly collimated jet originating from the TDE could be the source
\citep{BloomEtAl2011,ZaudererEtAl2011}, but the chances that it is pointing
exactly towards us should be very low. Additionally, the fraction of TDEs
producing a jet should also be negligible \citep{DeColleEtAl2012}. For the rest of
this article, we will hence not consider the jet model.

\begin{figure}
\includegraphics[width=84mm]{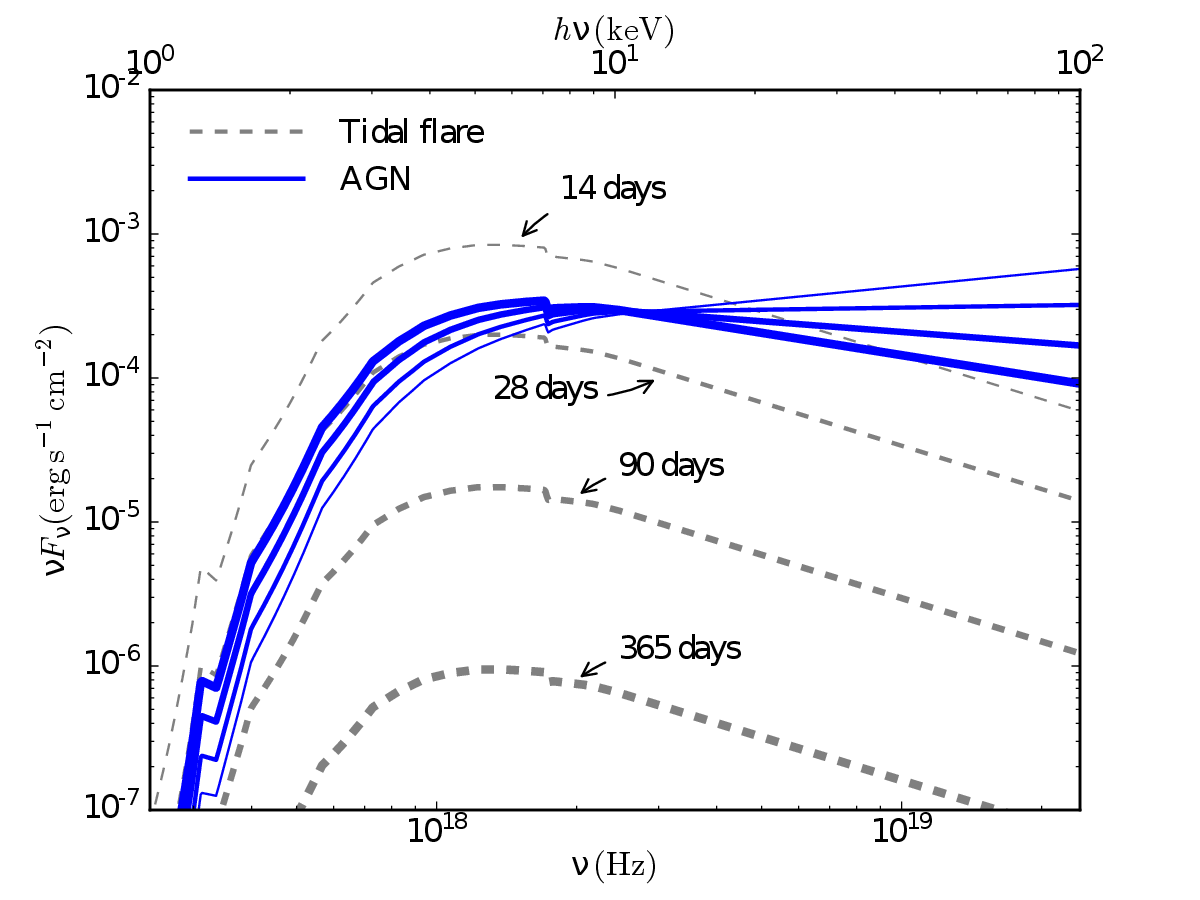}
\caption{
Spectral irradiances from the AGN-like and TDE scenarios.
The four solid lines with increasing thickness depict the irradiances
of the tidal disruption $(14,28,90,365)$ days after the disruption of a solar-type star.
The four dashed curves are for the AGN-like possibility, adopted from Figure~\ref{fig1}.
We apply the extinction law on the irradiance \citep{tan04}, so that the soft component is completely
absorbed and hence does not show in the figure.
\label{fig2}}
\end{figure}

In Figure~\ref{fig2} we compare the irradiances of the two separated scenarios:
The AGN-like and the TDE one.  These TDE candidates exhibit strong power-law
components with $2<\Gamma<5$ in the energy band $h\nu>2$ keV.  Observationally
we lack the cadence to catch the peak of the event, which should happen during
the first two weeks following the TDE.

Theoretically, the expected accretion during that time is so high, $L_{\rm
bol}>3\,L_{\rm Edd}$ according to our model, even approaching $10^2L_{\rm Edd}$
in the first couple of days, that there is not a well-established consensus on
our current accretion-disc models for it (see, for instance,
\cite{ulmer99,strubbe11,coughlin14,PiranEtAl2015,BegelmanVolonteri2017}).
I.e. we cannot convert the fallback rate of stellar debris into luminosity.
This is why we only display from day 14 onward in our figure.

In the figure we see that the TDE and the AGN-like models seem to be at similar
levels in the hard X-ray band, at least for a period of time not superior to one
month. However, we note that the structure of the star (its mass, and radius,
mostly), as well as the orbital features of the TDE can lead to significant
changes in the flux and timescales (see e.g. equation 13 in \cite{chen13}).

\section{Fluxes and fluences}

In the previous section we have seen that the integrated hard (i.e.  starting at
$\sim 2$ keV) X-ray flux is of about $10^{-3}~{\rm erg~s^{-1}~cm^{-2}}$,
comparable to that from the active sun. This is true for the AGN-like scenario
and also for the TDE one, but only for the first two weeks, since after that
period of time the flux drops significantly, up to three orders of magnitude
after the first year.

Besides irradiance, another quantity which is often used to evaluate the
lethality of ionizing radiation is the so-called fluence, i.e. the total amount
of radiation injected into one unit area. For this work, we consider fluence to
be the total integrated emitted energy in hard X-ray. We find that it is
$10^3~{\rm erg~cm^{-2}}$ for the TDE scenario, being conservative, since we
only consider a week of duration after the first two weeks. The AGN-like case
yields $3\times10^{9}~{\rm erg~cm^{-2}}$ where we have assumed a total duration
of $10^5$ years, to be consistent with the constraint from the kinematics of the
Fermi bubble \citep{ZubovasEtAl2011}.

The impact of hard X-ray flux and fluence on Earth has been discussed in the
related literature in the context of gamma-ray bursts (GRBs):

\noindent
(i) Flux: It has been shown by \cite{fishman88,inan99,mandea06} that in the past
30 years several distant (soft) gamma-ray bursts (GRBs) had induced similar
level of irradiation on Earth, with fluxes of about $10^{-3}~{\rm
erg~s^{-1}~cm^{-2}}$ at $h\nu>5$ keV, causing disturbances of the ionosphere.
The kind of flux that we consider in the AGN- and TDE-like scenarios could hence
potentially lead to a more serious disturbance of the ionosphere, because they
last longer, ranging from a few weeks to as long as $10^5$ years.

\noindent
(ii) Fluence: This is why probably it is more useful to look not at the flux,
but at the fluence - Possible nearby supernova numerical studies predict a hard
X-ray fluence in the range $(2-3)\times10^6~{\rm erg~cm^{-2}}$, which leads to
an ozone depletion of about $3\%$ \citep{melott11}. Fluences of about $10^8~{\rm
erg~cm^{-2}}$ lead to removal factors as high as $(20-30)\%$ \citep{ejzak07}.
While the TDE-like case is unlikely to reach these levels, as we have seen,
the AGN-like case exceeds this number by one order of magnitude.

\section{Discussion}

There is growing evidence that {\Sgr} must have been active in the past, more
specifically as recently as several Myr ago. This fact has motivated us to
study the irradiation from a recent active episode of {\Sgr} on Earth.

In this work we have estimated the flux on Earth emitted by such as an episode.
In particular, we have considered two different possibilities, a TDE and a
recent AGN-like activity. We show that for Earth, and out to Jupiter, the flux
from a violent X-ray activity originating at {\Sgr} is negligible compared to
the sun's. I.e., from about $10$ AU to the edge of the solar system, the
irradiation from \Sgr during its outburst predominates in the entire X-ray band
of $h\nu>1$ keV.  Most of the gas giants in the solar system, including their
moons, lie in this region, but the corresponding reaction of their atmosphere
has not been explored.

\begin{figure}
\includegraphics[width=84mm]{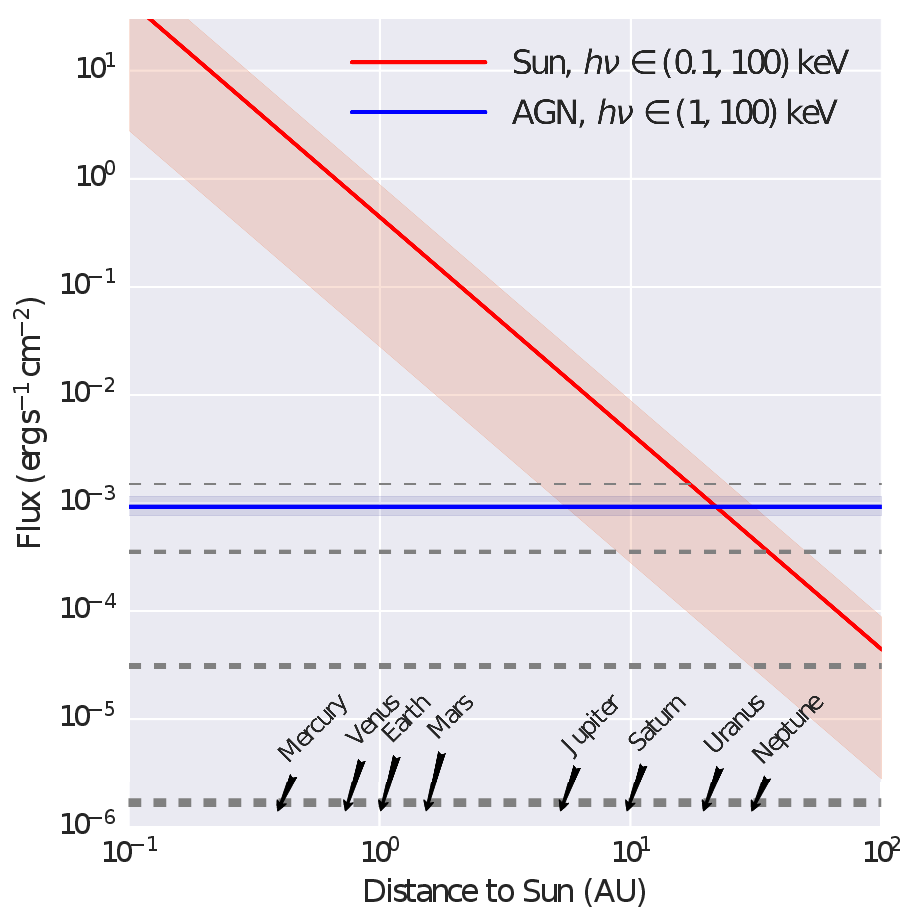}
\caption{
X-ray irradiances across the solar system for the TDE and AGN-like
scenarios. The TDE is shown as grey, dashed lines, following the same code
as in Fig.(\ref{fig2}) for the time duration. The AGN maximum and minimum
correspond to the upper and lower boundaries of the blue-shaded area, respectively,
and the thick blue line the average. The solar irradiance is displayed as a red-shaded
area. The upper- and lower limits correspond to the solar maximum and minimum, respectively,
and the thick red line the average of the two. We mark the position of the solar planets.
\label{fig3}}
\end{figure}

X-ray irradiation can also drive the chemical reactions in dense molecular
clouds \citep{krolik83}, and enhance the abundance of organic molecules in
protoplanetary discs \citep{teske11}. The implication is that the ancient
outbursts from \Sgr may have played an important role in shaping the habitable
environments on planets, especially on those exoplanets which are much closer
to the Galactic Centre than our solar system is.

\section*{Acknowledgements}

We thank Fukun Liu for his suggestion of using empirical SED templates, and Ran
Wang for providing references for SEDs. We thank Erlin Qiao and Jieying Liu for
the dedicated numerical simulations for the X-ray spectrum. We are grateful to
Emily Davidson for proofreading our manuscript.  This work was supported by the
National Key R\&D Program of China (2016YFA0400702) and the National Science
Foundation of China (11721303).  Pau Amaro Seoane thanks Marta Masini for her
outstanding and unique assistance and endurance during the last days in the
preparation of the manuscript in Beijing. He acknowledges support from the
Ram{\'o}n y Cajal Programme of the Ministry of Economy, Industry and
Competitiveness of Spain, as well as the COST Action GWverse CA16104.

\section*{References}

%\bibliography{Xray,aamnem99,biblio}

\end{document}